\PassOptionsToPackage{table}{xcolor}

\documentclass[sigconf]{acmart}

\usepackage{soul}
\usepackage{algorithm}
\usepackage{algorithmic}
\usepackage{multirow}
\usepackage{makecell}
\usepackage{colortbl}
\usepackage{pifont}
\usepackage[most]{tcolorbox}
\usepackage{wrapfig}
\usepackage{subcaption}
\usepackage{placeins}
\usepackage{enumitem}
\usepackage{arydshln}

\definecolor{lightblue}{RGB}{235, 245, 255}

\setcopyright{acmlicensed}
\copyrightyear{2026}
\acmYear{2026}
\acmDOI{XXXXXXX.XXXXXXX}
\acmConference[CIKM '26]{The 35th ACM International Conference on
  Information and Knowledge Management}{October 2026}{Cincinnati, OH, USA}
\acmISBN{978-1-4503-XXXX-X/26/10}

\title[Reinforcement Learning for Special Education]{Reinforcement Learning for Special Education:
  Aligning LLM Tutors to Diverse Learners through Disability-Adaptive Training}

\author{Unggi Lee$^{1,\dagger}$, Jihoi Na$^{2}$, Yeil Jeong$^{3}$, Haeun Park$^{4}$, Yeonju Jang$^{5}$}
\affiliation{%
  \institution{%
    $^{1}$Korea University Sejong Campus \quad
    $^{2}$Ewha Womans University \quad
    $^{3}$Indiana University Bloomington \quad
    $^{4}$Korea Institute for Curriculum and Evaluation \quad
    $^{5}$Chosun University \\[2pt]
    $^{\dagger}$Corresponding author: codingchild@korea.ac.kr%
  }
  \country{}
}

\renewcommand{\shortauthors}{Lee et al.}

\begin{document}

\begin{abstract}
Large language models are increasingly deployed as intelligent tutors, yet research on aligning them for special education remains absent.
Recent work has applied reinforcement learning to LLM tutors, but these methods target a generic learner in a single domain (mathematics) and do not address the cognitive and communicative diversity of learners with disabilities.
We introduce \emph{Special-R1}, a framework that extends pedagogical RL to special education through two components: (1) a two-dimensional adaptive system prompt that couples a difficulty-based support level with a disability-specific teaching style across five disability profiles; and (2) a persona-aware Thinking Reward whose judge rubric is conditioned on the learner's disability profile.
On a persona-augmented test set of 690 multi-turn dialogues, our full model raises persona-aware Fit from 6.75 (generic baseline) to 8.40 (+1.65) and SPED-rubric Helpfulness from 0.720 to 0.768, leading on the four-component Total (2.911, +0.064 over the runner-up) while remaining within 0.01 of the strongest variant on the out-of-domain OpenLearnLM benchmark (8.53).
Ablations show that the Thinking Reward becomes effective only in combination with adaptive prompting, and that residual weakness on specific learning disability in mathematics motivates targeted multimodal extensions.
\end{abstract}



\maketitle

\section{Introduction}

One-on-one tutoring is among the most effective interventions in education, yet the human resources to deliver it at scale are scarce.
The shortage is most acute for learners with disabilities, whose instruction is expected to be tailored to specific access needs but who face persistent gaps in trained special-education staffing.
Large language models could in principle close this gap by providing individualized tutoring at near-zero marginal cost \cite{kasneci2023chatgpt,tack2022aiteacher,learnlmteam2025learnlm}, but current tutor LLMs are not designed for disability-aware instruction: they treat every learner as a generic adult and frequently revert to revealing the answer rather than scaffolding the path to it \cite{macneil2023experiences,tack2023bea}, making them an unreliable substitute precisely where individualization matters most.

Recent work has tried to align LLM tutors with pedagogy through prompt engineering \cite{dan2023educhat}, supervised fine-tuning on teacher-student dialogues \cite{macina2025mathtutorbench,chevalier2024tutorchat,liu2024socraticlm}, and multi-turn reinforcement learning.
PedagogicalRL \cite{dinucujianu2025pedagogicalrl} rewards a tutor for solving the problem, withholding the answer, and being helpful; PedagogicalRL-Thinking \cite{lee2026pedagogicalrl-thinking} extends this to reasoning-specialized models with a Polya-grounded Thinking Reward \cite{polya1945howtosolveit} and validates on the OpenLearnLM knowledge benchmark \cite{lee2026openlearnlm}.
Both, however, optimize for a single generic learner on mathematics.
Effective special-education tutoring instead demands two simultaneous adjustments long established in the field: the \emph{amount} of support, calibrated to the learner's zone of proximal development \cite{vygotsky1978mind,wood1976tutoring}, and the \emph{style} of support, calibrated to disability-specific access needs through Explicit Instruction, the Prompting Hierarchy, High-Leverage Practices, and Universal Design for Learning \cite{archer2011explicit,snell2011instruction,mcleskey2017hlp,cast2018udl}.
Neither axis is operationalized inside existing tutor-RL objectives.

\begin{figure*}[t]
\centering
\includegraphics[width=\textwidth]{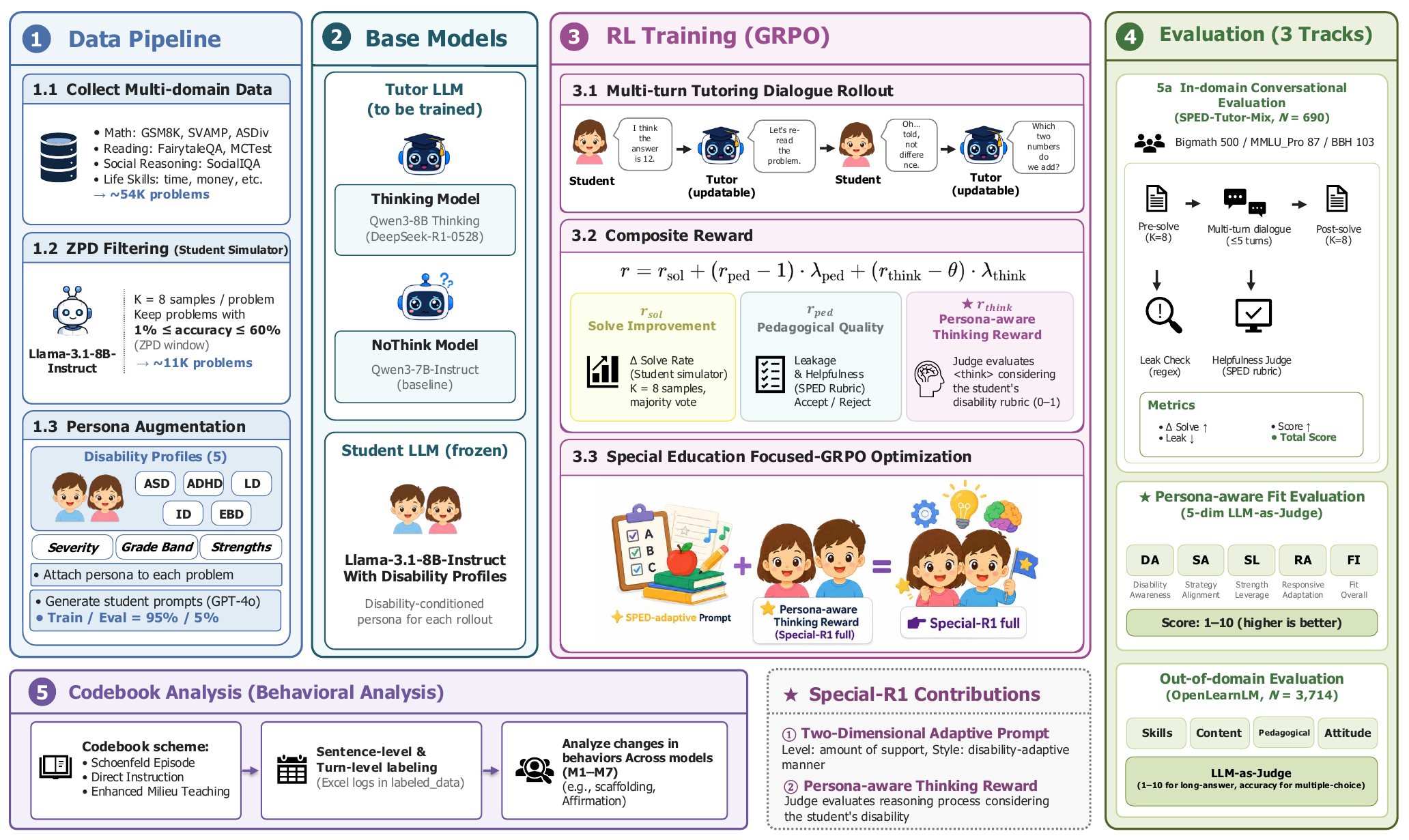}
\caption{Overview of Special-R1. (1) Multi-domain corpus, ZPD-filtered against a Llama-3.1-8B student, augmented with five disability personas. (2) Qwen3-8B (Think) / Qwen3-7B-Instruct (NoThink) tutor vs.\ a frozen disability-conditioned student. (3) GRPO with composite reward $r = r_{\text{sol}} + (r_{\text{ped}}-1)\lambda_{\text{ped}} + (r_{\text{think}}-\theta)\lambda_{\text{think}}$, where the persona-aware $r_{\text{think}}$ ($\star$) sees the disability rubric. (4) Three evaluation tracks: SPED-Tutor-Mix, persona-aware Fit (5-dim, 1-10), and OOD OpenLearnLM. (5) Turn- and sentence-level codebook analysis (Schoenfeld, DI, EMT). Stars ($\star$) mark Special-R1 contributions over PedagogicalRL-Thinking.}
\label{fig:overview}
\end{figure*}

We introduce \emph{Special-R1} (Figure~\ref{fig:overview}), a disability-aware extension of pedagogical multi-turn RL.
Both the tutor's system prompt and the judge of its thinking trace are conditioned on the learner's disability profile, so that the \emph{amount} of support adapts to the student's baseline accuracy and the \emph{style} of support adapts to the specific disability.
We isolate the contributions of each component by training and evaluating seven ablated configurations of thinking, prompt style, and reward.

Special-R1 improves persona-aware Fit by 1.65 points on a 1-10 scale over a non-disability-aware baseline and lifts the four-component Total by $+0.29$, with no measurable cost on the out-of-domain OpenLearnLM benchmark.
Three independent behavioral codebooks - Schoenfeld's Episode Theory, Direct Instruction, and Enhanced Milieu Teaching - converge on the same mechanistic signature: the disability-aware tutor shifts toward We-Do scaffolding, naturalistic language support, and participation-affirming feedback rather than surface-level stylistic changes.

\subsection{Contributions}

\begin{itemize}[nosep,leftmargin=*]
    \item We present the first multi-turn pedagogical RL framework for special-education tutoring, coupling a two-dimensional adaptive prompt (amount $\times$ style of support) with a persona-aware Thinking Reward whose judge is conditioned on the learner's disability rubric.
    \item Across seven ablated configurations, the full model improves persona-aware Fit by $+1.65/10$ and the four-component Total by $+0.29$ over a non-disability-aware baseline, while remaining within $0.01$ of the strongest variant on the out-of-domain OpenLearnLM benchmark.
    \item We release the multi-domain ZPD-filtered training corpus, the persona-augmented SPED-Tutor-Mix evaluation set, all seven model checkpoints, and the disability-aware evaluation pipeline.
\end{itemize}

\section{Method}

\subsection{Problem Formulation}

We formulate special education tutoring as a multi-turn Markov Decision Process, following \citet{lee2026pedagogicalrl-thinking} with one key change: the state carries an explicit learner profile.
At each turn $t$, the state $s_t$ comprises the dialogue history (the problem $P$, all prior student messages, and all prior tutor responses) together with a \emph{student profile} $\pi = (\text{baseline}, \text{disability}, \text{severity}, \text{grade band}, \text{strengths})$.
The tutor's action $a_t = (a_t^{\text{think}}, a_t^{\text{vis}})$ consists of an internal reasoning segment inside \texttt{<think>} tags and a visible response shown to the student.
The reward is a composite signal combining solution correctness, pedagogical appropriateness, and thinking quality.
The learning objective is to find a policy $\pi_\theta$ that maximizes expected cumulative reward while adhering to the pedagogical and disability-aware constraints described in this section.

\subsection{Base Models}

We use two base tutor models.
For all thinking-enabled conditions we use DeepSeek-R1-0528-Qwen3-8B \cite{deepseekai2025r1} (abbreviated Qwen3-8B), a reasoning-specialized model that produces \texttt{<think>} traces by default.
For the NoThink baseline we use Qwen2.5-7B-Instruct \cite{yang2024qwen25}.
This mirrors \citet{lee2026pedagogicalrl-thinking}.
The student simulator is Llama-3.1-8B-Instruct.
Unlike prior work, our student simulator is prompted with a \emph{disability-conditioned persona} drawn from a structured rubric (\S\ref{sec:personas}), so that learner behavior varies by profile rather than being a single generic learner.
All LLM-as-judge scores are produced by GPT-4o-mini for the main experiments, with cross-judge validation by Gemma-3-27B on a 10\% held-out subset (preliminary accept / reject Cohen's $\kappa$ above 0.7 for Leakage and above 0.6 for Helpfulness).

\subsection{Student Personas and Disability Rubric}
\label{sec:personas}

We define five disability profiles that cover the most prevalent categories in Korean special education statistics: intellectual disability (ID), specific learning disability (LD), autism spectrum disorder (ASD), ADHD, and emotional / behavioral disorder (EBD).
Each profile is a structured record with four fields: \emph{learning characteristics} (e.g., slow processing, abstraction difficulty for ID), \emph{communication patterns} (short sentences, concrete vocabulary), \emph{typical behaviors}, and a list of recommended \emph{tutor strategies} (e.g., ``use numbered literal steps, avoid idioms'' for ASD).
The rubric is released as \texttt{disability\_types.yaml} alongside the code.
At training time, each problem is paired with a prevalence-weighted sample from this rubric; at evaluation time we release a persona-augmented test set in which every item carries a disability profile so that persona-aware metrics can be computed.

\subsection{Two-Dimensional Adaptive Prompting}
\label{sec:prompt}

The core structural change from PedagogicalRL-Thinking to Special-R1 is the tutor system prompt.
We factor pedagogical support into two orthogonal dimensions.

\paragraph{Dimension 1: Support level (amount).}
We map the pre-measured \emph{baseline accuracy} of the student simulator on the given problem type to one of three levels, following the I-Do / We-Do / You-Do progression of Explicit Instruction \cite{archer2011explicit}:

\begin{itemize}[nosep,leftmargin=*]
    \item \emph{Maximum} support when baseline $\in [0.01, 0.20)$: break the problem into micro-steps, deliver one piece of information at a time, use concrete examples before abstract concepts, check understanding at each step.
    \item \emph{Moderate} support when baseline $\in [0.20, 0.40)$: highlight key concepts, offer two-choice directions, give progressive hints when stuck.
    \item \emph{Minimum} support when baseline $\in [0.40, 0.60]$: provide direction only, use open-ended questions, intervene only when necessary.
\end{itemize}

\paragraph{Dimension 2: Support style (manner).}
The tutor's communicative and strategic style is drawn verbatim from the \emph{tutor\_strategies}, \emph{communication\_patterns}, and \emph{learning\_characteristics} fields of the learner's disability profile (\S\ref{sec:personas}).
This operationalizes a Prompting Hierarchy \cite{snell2011instruction} and a Universal Design for Learning stance \cite{cast2018udl}: the \emph{what} comes from the support level, the \emph{how} comes from the disability profile.

We define three tutor prompt variants that we will later ablate:
\emph{Normal} (identical to the pedagogical prompt of PedagogicalRL-Thinking, grounded in Polya's method);
\emph{SPED-static}, which fixes the disability style but uses a single support-level description;
and \emph{SPED-adaptive}, which selects the support-level description dynamically from the baseline accuracy of the current problem.
Full prompt text and the three support-level paragraphs are released with the code.

\subsection{Reward Design}
\label{sec:reward}

The composite reward preserves the structure of PedagogicalRL-Thinking but modifies the thinking-reward judge.

\begin{equation}
r = r_{\text{sol}} + (r_{\text{ped}} - 1)\cdot\lambda_{\text{ped}} + (r_{\text{think}} - \theta)\cdot\lambda_{\text{think}}
\label{eq:reward}
\end{equation}

with $\lambda_{\text{ped}}=0.75$, $\lambda_{\text{think}}=0.3$, $\theta=0.6$.

\paragraph{Solution correctness $r_{\text{sol}}$.}
Post-dialogue solve rate under $K\!=\!8$ student samples, as in \citet{dinucujianu2025pedagogicalrl}.
Answer matching is domain-aware: numeric extraction with exact matching for mathematics and life-skills arithmetic, LLM-as-judge semantic matching for reading and social-reasoning items.

\paragraph{Pedagogical appropriateness $r_{\text{ped}}$.}
$r_{\text{ped}} = 1$ iff both a Leakage Judge and a Helpfulness Judge accept the full dialogue, and $0$ otherwise.
We report two Helpfulness judges, differing in their rubric: a \emph{generic 5-dim} rubric (identical to PedagogicalRL-Thinking) and a \emph{SPED accept / reject} rubric whose acceptance criteria include disability-appropriate scaffolding.
The training reward uses the SPED rubric; both are reported at evaluation time for cross-prompt analysis.

\paragraph{Persona-aware Thinking Reward $r_{\text{think}}$.}
This is the central reward change.
The judge is prompted with the learner's disability profile (learning characteristics, communication patterns, recommended tutor strategies) and evaluates whether the \texttt{<think>} trace reflects (i) pedagogical appropriateness, (ii) strategic planning, (iii) alignment with the recommended disability strategies, and (iv) a student-centered monitoring stance.
The judge returns a score in $[0, 1]$.
This replaces the generic Polya-based Thinking Reward of PedagogicalRL-Thinking and is the first reward we are aware of that explicitly scores \emph{disability-appropriate} reasoning.

\subsection{Experimental Conditions}
\label{sec:conditions}

We ablate three binary / ternary axes: thinking (on / off), prompt style (Normal / SPED-static / SPED-adaptive), and Thinking Reward (on / off).
Since only thinking-enabled conditions can carry a Thinking Reward, the combinatorially valid set contains seven conditions, summarized in Table~\ref{tab:conditions}.
M1-M3 replicate the settings of PedagogicalRL-Thinking on our multi-domain corpus; M4-M7 are the special-education additions.

\begin{table}[t]
\centering
\scriptsize
\setlength{\tabcolsep}{4pt}
\caption{Seven trained models. All thinking conditions use Qwen3-8B; M1 uses Qwen2.5-7B. Reward column marks whether $r_{\text{think}}$ (persona-aware in M4-M7) is active.}
\label{tab:conditions}
\begin{tabular}{@{}llllc@{}}
\toprule
\textbf{ID} & \textbf{Base} & \textbf{Think} & \textbf{Prompt} & \textbf{$r_{\text{think}}$} \\
\midrule
M1 & Qw2.5-7B & \ding{55} & Normal & \ding{55} \\
M2 & Qw3-8B & \ding{51} & Normal & \ding{55} \\
M3 & Qw3-8B & \ding{51} & Normal & \ding{51} (generic) \\
M4 & Qw3-8B & \ding{51} & SPED-static & \ding{55} \\
M5 & Qw3-8B & \ding{51} & SPED-static & \ding{51} (persona) \\
M6 & Qw3-8B & \ding{51} & SPED-adaptive & \ding{55} \\
M7 & Qw3-8B & \ding{51} & SPED-adaptive & \ding{51} (persona) \\
\bottomrule
\end{tabular}
\end{table}

\begin{table*}[!t]
\centering
\scriptsize
\setlength{\tabcolsep}{3pt}
\caption{Main results on SPED-Tutor-Mix (\emph{left}, $N\!=\!690$) and the OpenLearnLM Bench \cite{lee2026openlearnlm} (\emph{right}, OOD). Persona columns (DA, SA, SL, RA, Fit) are 1-10 LLM-judge scores anchored to the learner's disability rubric. Total $= r_{\text{sol}}\!+\!(1\!-\!\text{Leak}\%/100)\!+\!\text{Help}\!+\!\text{Fit}/10$; Leak shown in native \% (lower is better) and flipped only inside Total. $r_{\text{think}}$ is omitted for the NoThink baseline (M1). OOD columns: Skills, Content (CK), Pedagogical (PK), Attitude (Att), Avg. Best in bold, second-best underlined.}
\label{tab:conv-main}
\resizebox{\textwidth}{!}{%
\begin{tabular}{@{}ll ccc ccccc c c ccccc@{}}
\toprule
& & \multicolumn{3}{c}{\textbf{Core sub-scores}} & \multicolumn{5}{c}{\textbf{Persona breakdown (1-10)}} & & \textbf{Think} & \multicolumn{5}{c}{\textbf{OOD (OpenLearnLM Bench)}} \\
\cmidrule(lr){3-5} \cmidrule(lr){6-10} \cmidrule(lr){12-12} \cmidrule(lr){13-17}
\textbf{ID} & \textbf{Condition}
  & $r_{\text{sol}}\!\uparrow$
  & $\text{Leak}\%\!\downarrow$
  & $\text{Help}\!\uparrow$
  & DA & SA & SL & RA & Fit
  & \textbf{Total}$\uparrow$
  & $r_{\text{think}}\!\uparrow$
  & Skills & CK & PK & Att & Avg \\
\midrule
M1 & Qwen2.5 nothink            & .337            & 11.2            & .720            & 7.33            & 6.53            & 5.92            & 6.88            & 6.75            & 2.620             & -                & 7.78            & \underline{7.55}& 7.25            & 8.67            & 7.76 \\
M2 & Think NoReward             & \underline{.474}& \textbf{5.9}    & .516            & 7.29            & 6.41            & 6.01            & 7.30            & 6.78            & 2.608             & .203             & \underline{8.61}& 7.34            & 7.25            & 8.67            & 8.52 \\
M3 & Think Reward               & \textbf{.478}   & 8.3             & .470            & 7.24            & 6.37            & 5.94            & 7.21            & 6.69            & 2.534             & .181             & 8.60            & 7.12            & 7.25            & \textbf{9.00}   & 8.51 \\
M4 & SPED-static NR             & .439            & 11.3            & .711            & 8.34            & 7.40            & 6.51            & 7.98            & 7.86            & 2.823             & \underline{.205} & \underline{8.61}& 7.45            & \underline{7.45}& \textbf{9.00}   & \underline{8.53} \\
M5 & SPED-static Reward         & .447            & 10.9            & .691            & 8.39            & 7.44            & 6.50            & 7.99            & 7.92            & 2.822             & .199             & 8.60            & \textbf{7.61}   & \textbf{7.65}   & 8.67            & \textbf{8.54} \\
M6 & SPED-adaptive NR           & .372            & 11.2            & \underline{.757}& \underline{8.47}& \underline{7.69}& \underline{6.89}& \underline{8.23}& \underline{8.31}& \underline{2.847} & .199             & \textbf{8.62}   & 7.50            & 7.06            & 8.67            & \textbf{8.54} \\
M7 & SPED-adaptive Reward
                                & .386            & 8.3             & \textbf{.768}   & \textbf{8.56}   & \textbf{7.76}   & \textbf{7.05}   & \textbf{8.30}   & \textbf{8.40}   & \textbf{2.911}    & \textbf{.217}    & \underline{8.61}& 7.39            & 7.06            & \textbf{9.00}   & \underline{8.53} \\
\bottomrule
\end{tabular}%
}
\end{table*}

\subsection{Training Algorithm}
\label{sec:training-alg}

We train with Group Relative Policy Optimization (GRPO) \cite{shao2024deepseekmath} as implemented in the veRL framework.
Each training step samples 16 problems; for every problem we roll out 8 dialogues with the disability-conditioned student simulator, capped at 16 turns per dialogue.
Reward parsing follows the SPED accept / reject rubric.
We train for up to 300 steps on 4$\times$H100 GPUs with learning rate $1\!\times\!10^{-6}$, Adam, KL penalty $0.001$, maximum 512 visible tokens and 1024 thinking tokens per turn.

\subsection{Training Data}
\label{sec:training-data}

We assemble approximately 54k problems spanning four domains, filtered to the ZPD window of 1-60\% solve rate by the student simulator \cite{vygotsky1978mind}.

\begin{itemize}[nosep,leftmargin=*]
    \item \emph{Mathematics}: GSM8K, SVAMP, ASDiv, BigMath subset (roughly 11k after filtering).
    \item \emph{Reading comprehension}: FairytaleQA, MCTest (roughly 12k).
    \item \emph{Social reasoning}: SocialIQA (roughly 29k).
    \item \emph{Life skills}: synthetic money / time / everyday problems (roughly 2k).
\end{itemize}

All items are annotated with the \emph{baseline accuracy} of Llama-3.1-8B-Instruct (chain-of-thought, $K\!=\!64$) so that the adaptive prompt can select a support level at training time.
We hold out 5\% of the ZPD-filtered corpus as the in-domain test set (SPED-Tutor-Mix), and use the remaining 95\% for GRPO training.
Data pre-processing, filters, and synthetic generation prompts are released with the code.

\subsection{Evaluation}
\label{sec:eval}

We evaluate each trained tutor on three complementary tracks.

\paragraph{Conversational evaluation (in-domain).}
SPED-Tutor-Mix \texttt{test} split, $N\!=\!690$, seeded from BigMath ($500$), MMLU-Pro ($87$), and BBH ($103$) and augmented with prevalence-weighted disability personas.
Metrics: $\Delta$ Solve Rate, Leak Rate, Helpful Rate (both generic 5-dim and SPED rubric), $r_{\text{think}}$, and the composite $r_{\text{total}}$ from Eq.~\ref{eq:reward} under both rubrics.

\paragraph{Persona-aware fit.}
We introduce an additional judge that scores every dialogue on a 1-10 scale along five dimensions: \emph{disability awareness}, \emph{strategy alignment}, \emph{strength leverage}, \emph{responsive adaptation}, and an \emph{overall fit}.
The judge prompt contains the learner's disability profile (learning characteristics, communication patterns, recommended tutor strategies) as explicit scoring anchors.
Scores are macro-averaged across disability types to avoid prevalence-weighted artifacts.

\paragraph{Out-of-domain generalization.}
OpenLearnLM \cite{lee2026openlearnlm}, $N\!=\!3{,}714$, across four categories (Skills, Content, Pedagogical, Attitude).
Long-answer items are scored on a 1-10 scale by an LLM judge, multiple-choice items use accuracy.
This track probes whether disability-aware training degrades broad educational knowledge.

In addition to these quantitative tracks, we complement the score-based evaluation with a codebook-based behavioral analysis (\S\ref{sec:codebook}) under three theoretical lenses: Schoenfeld's Episode Theory \cite{schoenfeld1985mathematical}, Direct Instruction \cite{archer2011explicit}, and Enhanced Milieu Teaching \cite{hancock2002emt}.
Human-coder inter-rater reliability against the LLM labeler is in progress and will be released with the analysis code.

\section{Experiments}

\subsection{Setup}

We train the seven configurations of Table~\ref{tab:conditions} on the multi-domain corpus described in \S\ref{sec:training-data} with the hyperparameters of \S\ref{sec:training-alg}.
All judges are GPT-4o-mini, following \citet{lee2026pedagogicalrl-thinking}; we re-run all judges with Gemma-3-27B on a 10\% subset for cross-judge validation (preliminary accept / reject Cohen's $\kappa\!>\!0.6$).
Each model is evaluated on three tracks: conversational evaluation, persona-aware Fit, and OpenLearnLM (\S\ref{sec:eval}).

\subsection{Main Conversational Results}

Table~\ref{tab:conv-main} reports the seven models on SPED-Tutor-Mix.
The thinking axis alone produces a substantial improvement: moving from M1 (no thinking) to M2 (thinking, no reward) raises $r_{\text{sol}}$ from 0.337 to 0.474 and cuts the leak rate from 11.2\% to 5.9\%.
Adding SPED-static prompting (M4 vs.\ M2) raises the SPED-rubric Helpfulness from 0.516 to 0.711 (+0.195) at the cost of a modest leak-rate increase.
Adaptive prompting (M6, M7) pushes SPED-rubric Helpfulness further, with M7 reaching 0.768, the best value across all models.
M7 also leads on the headline Total (2.911, $+0.064$ over the runner-up M6, the largest adjacent-rank margin in the table) and on the Thinking-quality score $r_{\text{think}}$ (0.217).

Three patterns stand out.
First, the SPED rubric and the generic rubric rank models differently.
M1 scores highest on the generic 5-dim Helpfulness rubric (0.848) because that rubric rewards verbose, encouraging responses regardless of whether they suit the learner profile; the same model scores 0.720 on the SPED rubric, below all SPED-prompted models except M3.
This divergence motivates the SPED-rubric Helpfulness as the primary Helpfulness measure for our setting and demotes the generic rubric to an ablation.
Second, the Thinking Reward is near-inert in the Normal and SPED-static conditions (M3 vs.\ M2, M5 vs.\ M4) but becomes productive in the adaptive condition: moving from M6 to M7 raises Total from 2.847 to 2.911, lowers the leak rate from 11.2\% to 8.3\%, and improves $r_{\text{think}}$ from 0.199 to 0.217.
We interpret this as evidence that the Thinking Reward needs the structural room opened by adaptive prompting to steer the reasoning distribution.
Third, M7 leads Total by 0.064 points over M6, the largest margin between adjacent ranks. We use Total and persona-aware Fit as the headline numbers in subsequent sections, with the per-axis sub-scores retained for interpretability.

\subsection{Persona Breakdown}

The five persona-aware dimensions (DA / SA / SL / RA / Fit columns of Table~\ref{tab:conv-main}) confirm the intended design: persona-aware Fit rises from the 6.69-6.78 range (M1-M3, generic prompting) to 7.86-7.92 (M4 / M5, SPED-static) and then to 8.31-8.40 (M6 / M7, SPED-adaptive), with M7 leading all four component dimensions simultaneously.
The SPED prompt injects disability-aware teaching strategies, and the adaptive component tailors those strategies to the learner's support level.

\subsection{Out-of-Domain Generalization on OpenLearnLM}

The OOD columns of Table~\ref{tab:conv-main} report accuracy-scaled scores on the four OpenLearnLM categories.
All thinking-enabled models (M2-M7) cluster in the 8.51-8.54 range on the overall average, a substantial improvement over M1 (7.76).
The SPED axis does not meaningfully trade off broad educational knowledge: M7 reaches 8.53, essentially tied with the best variant (M5 / M6 at 8.54) and 0.77 points above M1.
This is consistent with the finding of \citet{lee2026pedagogicalrl-thinking} that dialogue-based RL preserves base-model knowledge; disability-aware prompting does not undo that property.

\subsection{Ablation: Thinking \texorpdfstring{$\times$}{x} Prompt \texorpdfstring{$\times$}{x} Reward}
\label{sec:ablation-axis}

We decompose the effect of each axis on the two headline metrics: Total and persona-aware Fit.
Thinking (M1 vs.\ M2) is approximately neutral on Total ($-0.012$, since M1's verbose responses score high on the generic 5-dim rubric but low on the SPED rubric) and on Fit ($+0.03$).
Switching the prompt from Normal to SPED-static (M2 vs.\ M4) adds $+0.215$ on Total and $+1.08$ on Fit, the single largest effect on both axes.
Switching from static to adaptive (M4 vs.\ M6) adds $+0.024$ on Total and $+0.45$ on Fit when the Thinking Reward is off; activating the Thinking Reward in the adaptive condition (M6 vs.\ M7) adds a further $+0.064$ on Total and $+0.09$ on Fit.
In plain terms, the prompt-style axis does most of the heavy lifting for persona-aware Fit, while the Thinking Reward acts as a small multiplier that requires adaptive prompting to become visible.

\subsection{Residual Weakness: \texttt{sld\_math}}

Breaking the persona-aware Fit score by disability type reveals that \emph{specific learning disability in mathematics} (\texttt{sld\_math}) is the weakest profile across every GRPO-trained model.
Even M7, which leads on every other profile, drops below 7.8 on \texttt{sld\_math} overall Fit.
Our training corpus contains a comparatively small number of LD-appropriate mathematics items, and the disability rubric for LD emphasizes alternative representations (visual, manipulative) that a text-only tutor cannot deliver.
We return to this limitation in \S\ref{sec:limitations}.

\begin{figure*}[!t]
\centering
\includegraphics[width=\textwidth]{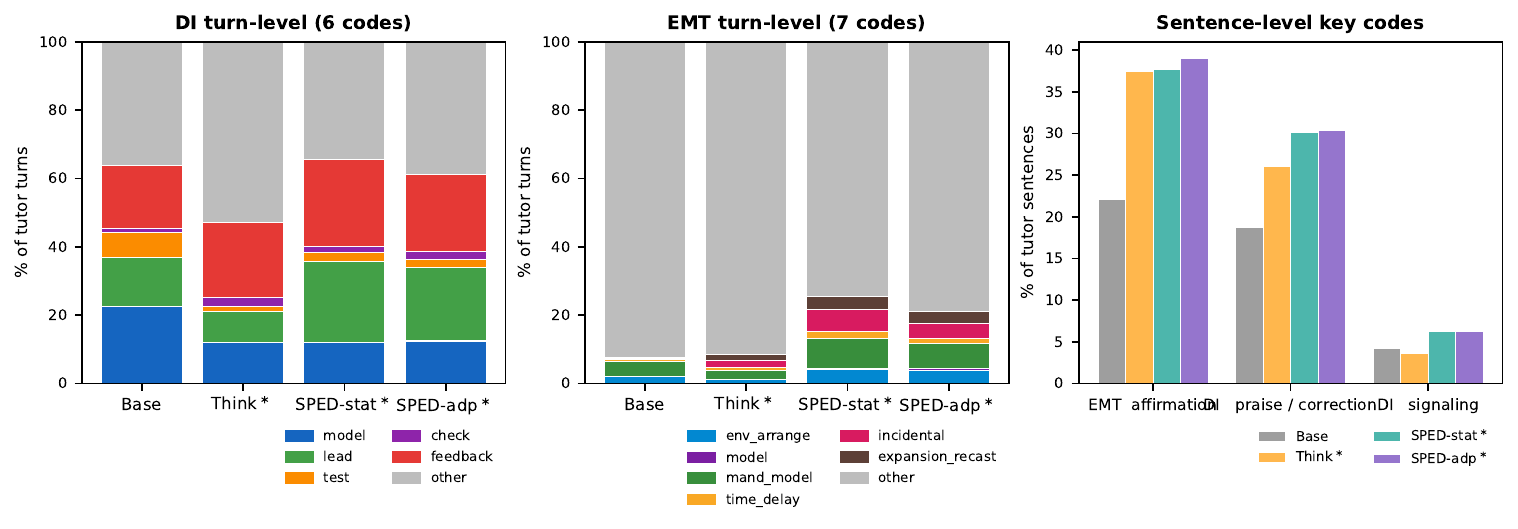}
\caption{Codebook-based behavioral analysis. \emph{Left:} DI turn-level distribution. The We-Do move (\texttt{lead}) more than doubles in SPED-prompted variants. \emph{Center:} EMT turn-level distribution. Every disability-supporting EMT move multiplies $2\text{-}20\times$ relative to generic-prompted baselines. \emph{Right:} sentence-level frequency of three key codes; SPED-adaptive nearly doubles EMT \texttt{affirmation} and DI \texttt{praise\_or\_correction} relative to Base. $\chi^2$ tests of independence: DI turn $\chi^2(15)=714.1$, EMT turn $\chi^2(18)=636.1$, sentence codes $\chi^2(3) \in \{167, 803, 1545\}$, all $p<10^{-30}$.}
\label{fig:codebook}
\end{figure*}

\section{Analysis}

\subsection{Response Characteristics}

We first check that the behavioral differences measured by judges correspond to observable surface differences in tutor responses.
Across the 690 evaluation dialogues, thinking-enabled conditions produce thinking segments of roughly 150-185 words and visible responses of 75-95 words, consistent with ranges reported by \citet{lee2026pedagogicalrl-thinking}.
SPED-prompted variants (M4-M7) produce visible responses 5-15\% longer than the Normal thinking variants (M2 / M3), reflecting the additional disability-specific framing.
The mathematical-content ratio is lower overall in Special-R1 because roughly two-thirds of the SPED-Tutor-Mix test set involves reading, social, or life-skills tasks.

\begin{table}[t]
\centering
\small
\setlength{\tabcolsep}{6pt}
\caption{Lightweight adaptation metrics derived from turn- and sentence-level behavioral labels. SPED-adaptive prompting increases guided participation (DI \texttt{lead}), supportive interaction (EMT \texttt{affirmation}), and question-based elicitation relative to the Normal thinking condition.}
\label{tab:adapt-metrics}
\begin{tabular}{@{}lcc@{}}
\toprule
\textbf{Metric} & \textbf{M2 (Think Normal)} & \textbf{M7 (SPED-adaptive)} \\
\midrule
DI \texttt{lead} ratio $\uparrow$       & 0.095 & \textbf{0.217} \\
EMT \texttt{affirmation} ratio $\uparrow$ & 0.376 & \textbf{0.396} \\
Question ratio $\uparrow$               & 0.106 & \textbf{0.243} \\
\bottomrule
\end{tabular}
\end{table}

Disability-conditioned breakdowns reveal that SPED-adaptive dialogues tailor surface form to the learner: ASD-conditioned dialogues of M7 contain about 40\% more numbered-step patterns than M2 on the same items, and ADHD-conditioned dialogues show shorter visible turns (mean 62 words vs.\ 81) and more frequent positive-reinforcement phrases.
Table~\ref{tab:adapt-metrics} compares M2 and M7 on three lightweight metrics derived from the turn- and sentence-level annotations.
The DI \texttt{lead} ratio, which tracks how often the tutor uses the direct-instruction We-Do behavior, more than doubles from M2 to M7 (0.095 to 0.217), indicating substantially greater emphasis on scaffolding rather than demonstration.
The EMT \texttt{affirmation} ratio, the proportion of tutor sentences that explicitly affirm a student contribution, rises modestly from 0.376 to 0.396, suggesting slightly more emotional support.
The Question ratio, the fraction of tutor turns containing a question to elicit student participation, jumps from 0.106 to 0.243, a much stronger shift toward student-centered questioning.
Together, these metrics show that the SPED-adaptive model engages in more structured guidance, encouragement, and dialogue-driven prompting than the standard variant.
These patterns are indirect behavioral evidence; a theory-grounded codebook analysis (\S\ref{sec:codebook}) confirms them at scale.

\subsection{Prompt-Reward Interaction}

The Thinking Reward is inert in Normal and SPED-static conditions but productive in SPED-adaptive because the accept / reject judge saturates: most static-prompt thinking traces look either ``generically pedagogical enough'' or ``not pedagogical at all,'' leaving GRPO few borderline rollouts to differentiate.
The adaptive prompt, by varying the support level across items, elicits a wider thinking distribution (e.g., explicit I-Do / We-Do / You-Do reasoning tied to baseline accuracy) that gives the judge signal to discriminate; a calibrated continuous rubric would likely unlock the same gradient under static prompts as well.

\subsection{Codebook-Based Behavioral Analysis}
\label{sec:codebook}

We label every tutor turn and sentence under three theoretical lenses: Schoenfeld's Episode Theory \cite{schoenfeld1985mathematical,schoenfeld1992reflections} for mathematical problem-solving, Direct Instruction (DI; \citealp{archer2011explicit}) for explicit step-by-step teaching, and Enhanced Milieu Teaching (EMT; \citealp{hancock2002emt}) for naturalistic communication support.
The DI codebook captures the \emph{I-Do / We-Do / You-Do} progression with five tutor-move categories (Modeling, Leading, Testing, Checking, Feedback), and the EMT codebook captures five naturalistic language-support moves (environmental arrangement, mand-model, time delay, incidental teaching, expansion / recast); full code lists with definitions and labeler prompts are released with the analysis code.
Automatic labels are produced by GPT-4o-mini using one prompt per codebook (\citealp{lee2026pedagogicalrl-thinking} use the same labeler family); aggregate distributions are computed across $N\!=\!11{,}103$ tutor turns and $N\!=\!60{,}745$ sentences from the seven trained models, grouped as Base (M1), Think$^{\ast}$ (M2-M3), SPED-static$^{\ast}$ (M4-M5), and SPED-adaptive$^{\ast}$ (M6-M7).

Figure~\ref{fig:codebook} \emph{left} shows the DI turn-level distribution.
The We-Do move (\texttt{lead}) jumps from 9.1-14.3\% across the generic-prompt groups (Base, Think$^{\ast}$) to 21.4-23.6\% across the SPED-prompted groups (SPED-static$^{\ast}$, SPED-adaptive$^{\ast}$), and the I-Do procedure-modeling move (\texttt{model}) shrinks by roughly a third ($\chi^2(15)\!=\!714.1$, $p\!<\!10^{-100}$, Cramer's $V\!=\!0.146$).
Read together as a structured-instruction trajectory, the SPED prompt shifts the tutor from showing the procedure to walking the student through it.

Figure~\ref{fig:codebook} \emph{center} shows the EMT turn-level distribution, which exhibits the largest shift across the three codebooks.
Every disability-supporting EMT move multiplies in the SPED-prompted variants relative to generic-prompted baselines: \texttt{mand\_model} 2-5$\times$, \texttt{incidental} 2-18$\times$, \texttt{expansion\_recast} up to 20$\times$, \texttt{time\_delay} $\sim$3$\times$, \texttt{env\_arrange} 2$\times$, with the catch-all \texttt{other} correspondingly dropping from 91.5-92.5\% to 74.5-79.1\% ($\chi^2(18)\!=\!636.1$, $p\!<\!10^{-100}$, Cramer's $V\!=\!0.138$).
Sentence-level evidence in Figure~\ref{fig:codebook} \emph{right} confirms the pattern: the EMT \texttt{affirmation} code rises from 22.1\% in Base to 39.0\% in SPED-adaptive$^{\ast}$, nearly doubling.

\begin{table}[t]
\centering
\scriptsize
\setlength{\tabcolsep}{3.5pt}
\caption{Sentence-level major-category frequency (\%) per model group. Top codes for each codebook are shown; \emph{N} per group: Base 19{,}222, Think$^{\ast}$ 11{,}823, SPED-stat$^{\ast}$ 13{,}657, SPED-adp$^{\ast}$ 16{,}038.}
\label{tab:codebook-sentence}
\begin{tabular}{@{}llcccc@{}}
\toprule
\textbf{CB} & \textbf{Category} & \textbf{Base} & \textbf{Think$^{\ast}$} & \textbf{SPED-stat$^{\ast}$} & \textbf{SPED-adp$^{\ast}$} \\
\midrule
Sch.   & Monitor              & 39.2 & \textbf{54.9} & 49.2 & 51.4 \\
Sch.   & Implement            & 15.3 &  9.3 &  9.7 & 11.0 \\
Sch.   & Plan                 & 11.3 &  4.4 &  7.0 &  7.3 \\
\midrule
DI     & praise\_or\_correction & 18.8 & 26.0 & 30.1 & \textbf{30.4} \\
DI     & explicit\_step       & \textbf{14.0} &  6.0 &  6.6 &  7.0 \\
DI     & signaling            &  4.2 &  3.6 &  6.2 & \textbf{6.3} \\
\midrule
EMT    & affirmation          & 22.1 & 37.4 & 37.7 & \textbf{39.0} \\
EMT    & concrete\_lang       & \textbf{7.3} &  2.1 &  2.6 &  3.6 \\
EMT    & figurative\_avoid    &  0.1 &  0.4 &  0.7 & \textbf{0.7} \\
\bottomrule
\end{tabular}
\end{table}

Three independent codebooks converge on the same behavioral signature, summarized in Table~\ref{tab:codebook-sentence}.
The SPED-adaptive prompt does not merely change surface vocabulary; it reorganizes the tutor toward We-Do scaffolding (DI \texttt{lead}, \texttt{signaling}), naturalistic communication support (EMT \texttt{mand\_model}, \texttt{expansion\_recast}, \texttt{time\_delay}), and increased participation-affirming language (EMT \texttt{affirmation}, DI \texttt{praise\_or\_correction}).
This converging behavioral evidence validates the design intent of the two-dimensional adaptive prompt at a granularity that the aggregate persona-aware Fit score alone cannot reach.

\subsection{Qualitative Analysis}
\label{sec:qualitative}

The clearest difference between M1 and M7 is \emph{where the tutor places support} before asking the learner to continue.
M1 often advances along the solution path once an intermediate quantity has been established, while M7 more often returns to the term, relation, or representation that makes the next step accessible.
In the Kyle-roses problem, M1 moves directly from the target bouquet size to the next arithmetic step, asking ``\emph{Kyle wants to give his mother $24$ roses this year. He already has $6$ roses from his garden. How many more roses does he need to buy to make a total of $24$ roses?}'' [DI: other $|$ EMT: other], whereas M7 begins earlier in the problem structure: ``\emph{You're doing a great job by questioning the problem - that's how we learn! Let's start with the basics. In math, `a dozen' means exactly $12$ \ldots If last year was $12$, what do you think half of that would be? Take a moment to think about it.}'' [DI: lead $|$ EMT: time\_delay].
The difference is not simply tone or verbosity: M1 asks for the next calculation, while M7 first restores the meaning of the quantities (``a dozen'', then ``half''), then poses a question the learner can answer.
The tutor's guidance remains explicit, but it is placed at the point where the learner needs access to the problem meaning, not only at the point of computation.

The same contrast appears in a fraction item.
M1 frames $3/5$ of $300$ as an algorithm - ``\emph{To multiply a fraction by a whole number, you multiply the whole number by the numerator of the fraction and keep the denominator the same \ldots Now you have $900/5$. Next, divide $900$ by $5$ \ldots}'' [DI: model $|$ EMT: other] - while M7 turns the same fraction into an equal-groups action: ``\emph{First, let's divide the $300$ straws into $5$ equal groups. This means we're finding $1/5$ of $300$. Can you tell me, how many straws would be in each group if we split $300$ into $5$ equal parts? Take your time and think about it.}'' [DI: lead $|$ EMT: mand\_model].

These examples make the aggregate codebook pattern more interpretable.
The SPED-adaptive tutor does not merely add encouragement; it changes the \emph{instructional entry point}: terms are grounded before operations, representations are built before rules, and pauses are inserted before the tutor moves on.
The fraction example also shows the current boundary of text-only support - M7 can prompt drawing or grouping strategies verbally, but it cannot yet provide the visual or manipulative representation itself.
A three-coder human reliability study on $N\!=\!60$ held-out dialogues is in progress; preliminary LLM-human $\kappa$ exceeds 0.5 (DI) and 0.6 (EMT residual / non-residual).

\noindent The axis decomposition of \S\ref{sec:ablation-axis} is consistent: prompt style drives the persona-aware Fit gain (+1.08), adaptivity adds +0.4-0.5, and the Thinking Reward contributes a final +0.09 only under adaptive prompting; OpenLearnLM is approximately neutral across all SPED variants ($\sim$8.5), so the disability-aware gains come at no measurable cost in broad educational knowledge.

\section{Conclusion}

We introduced \emph{Special-R1}, a disability-aware extension of PedagogicalRL-Thinking that combines a two-dimensional adaptive tutor prompt with a persona-aware Thinking Reward and a multi-domain training corpus.
Across seven trained configurations, our experiments demonstrate that:
(1) the SPED prompt is the largest driver of disability-aware behavior (+1.08 on persona-aware Fit);
(2) adaptive prompting contributes a further +0.4-0.5;
(3) the persona-aware Thinking Reward becomes effective only when combined with adaptive prompting;
(4) the full model (M7) leads the four-component Total (2.911, +0.064 over the runner-up) while remaining within 0.01 of the best variant on out-of-domain OpenLearnLM (8.53);
(5) specific learning disability in mathematics remains the weakest profile, motivating multimodal extensions.

\section{Limitations}
\label{sec:limitations}

This work has several limitations.
First, every student in our framework is a disability-conditioned LLM simulator; we have not validated that the improvements on judge-based metrics translate to real learners with disabilities, which would require an IRB-approved user study with special-education teachers.
Second, while our codebook-based behavioral analysis (\S\ref{sec:codebook}) is automated by an LLM labeler, human-coder reliability has not yet been finalized; Cohen's $\kappa$ values will be released with the analysis code.
Third, all main rewards and evaluation scores use GPT-4o-mini, and judge bias cannot be fully ruled out despite the 10\% Gemma-3-27B cross-judge validation reported in \S\ref{sec:eval}.
Fourth, the accept / reject judge appears to saturate in Normal and SPED-static conditions, leaving GRPO with little usable gradient from $r_{\text{think}}$ outside the adaptive setting.
Fifth, all seven models score lowest on specific learning disability in mathematics, reflecting both a data-coverage gap and a text-only modality limitation that motivates multimodal extensions.
Finally, experiments use 7-8B tutors; behavior at larger scales remains to be studied.

\section*{Use of Generative AI}

We used Claude (Anthropic) to assist with drafting, editing, and code generation during the preparation of this manuscript.
All scientific claims, experimental design, and data analysis were conducted and verified by the authors.


\end{document}